\documentclass[11pt, a4paper,prd,superscriptaddress,nofootinbib,aps,preprint]{revtex4-2}
\usepackage[utf8]{inputenc}
\usepackage{%
  mathtools,
  amsfonts, 
  physics,
  xcolor,
  hyperref,
}

\usepackage{overpic,}

\hypersetup{pageanchor=false}

\definecolor{CUDred}{RGB}{255,75,0}    
\definecolor{CUDgreen}{RGB}{3,175,122} 
\definecolor{CUDblue}{RGB}{0,90,255}   

\hypersetup{
  colorlinks        = true,
  bookmarksopen     = true,
  bookmarksnumbered = true,
  allcolors         = CUDred
}

\usepackage[T1]{fontenc}
\usepackage{mlmodern}
\renewcommand{\frac}[2]{\genfrac{}{}{0.6pt}{}{#1}{#2}}

\newcommand{\Emph}[1]{\emph{\textcolor{CUDblue}{#1}}}
\newcommand{\Rev}[1]{#1}

\newcommand{\Tensor}[1]{\boldsymbol{#1}}

\newcommand{\ManifoldSymbol}{M}
\newcommand{\ContactFormSymbol}{\eta}
\newcommand{\ReebVectorSymbol}{\xi}
\newcommand{\PhiTensorSymbol}{\varphi}
\newcommand{\CompatibleMetricSymbol}{h}

\newcommand{\FrameFieldSymbol}{e}
\newcommand{\CoframeFieldSymbol}{\omega}
\newcommand{\BaseSpaceSymbol}{B}
\newcommand{\FourMetricSymbol}{g}
\newcommand{\NullFrameFieldRealLineSymbol}{k}

\newcommand{\Manifold}{\mathcal{\ManifoldSymbol}}
\newcommand{\ThreeMani}{\Manifold^3}
\newcommand{\ContMani}{\Manifold}
\newcommand{\BaseSpace}{\mathcal{\BaseSpaceSymbol}}

\newcommand{\ContForm}{\Tensor{\ContactFormSymbol}}

\newcommand{\dContForm}{\dd \ContForm}
\newcommand{\dContFormComp}{\dd \ContactFormSymbol}

\newcommand{\ReebVect}{\Tensor{\ReebVectorSymbol}}

\newcommand{\PhiTens}{\Tensor{\PhiTensorSymbol}}
\newcommand{\PhiTensComp}{\PhiTensorSymbol}

\newcommand{\CompatibleMetric}{\Tensor{\CompatibleMetricSymbol}}

\newcommand{\ContMet}{\CompatibleMetric}
\newcommand{\ContMetComp}{\CompatibleMetricSymbol}

\newcommand{\FrmFld}{\Tensor{\FrameFieldSymbol}}

\newcommand{\CofFld}{\Tensor{\CoframeFieldSymbol}}

\newcommand{\LieDeriv}{\mathcal{L}}
\newcommand{\IntProd}{\iota}
\newcommand{\EmptySlot}{\,\cdot\,}
\newcommand{\RicTens}{\Tensor{\mathrm{Ric}}}
\newcommand{\R}{\mathbb{R}}

\newcommand{\BaseRicScal}{R_{\BaseSpace}}

\newcommand{\BaseMet}{\ContMet_{\BaseSpace}}

\newcommand{\EinEq}{the Einstein equation}
\newcommand{\Nondegenerate}{nondegenerate}
\newcommand{\Nonvanishing}{non-vanishing}
\newcommand{\Ppwaves}{\textit{pp}-waves}

\newcommand{\FourMet}{\Tensor{\FourMetricSymbol}}
\newcommand{\FourRicTens}{\RicTens}

\newcommand{\NullFrmFldRL}{\Tensor{\NullFrameFieldRealLineSymbol}}
\newcommand{\NullFrmFldReeb}{\ReebVect}

\newcommand{\ThreeRicTens}{\RicTens_{\ThreeMani}}
\newcommand{\ThreeContraRicTens}{\prescript{\sharp}{}{\RicTens_{\ThreeMani}}}

\newcommand{\FourContraRicTens}{\prescript{\sharp}{}{\RicTens}}

\newcommand{\EinTens}{\prescript{\sharp}{}{\Tensor{G}}}

\newcommand{\EneMomTens}[2][]{\prescript{\sharp}{}{\Tensor{T}}^{#1}_{\text{#2}}}

\newcommand{\EneDensNullDust}{\Phi^2}

\newcommand{\LablCosmicString}{\text{string}}
\newcommand{\LablCosmicStrings}{\text{strings}}
\newcommand{\LablNullDust}{\text{null dust}}
\newcommand{\LablEinHil}{\text{EH}}
\newcommand{\LablMin}{\text{min}}
\newcommand{\LablMax}{\text{max}}

\newcommand{\Action}[1]{S_{\text{#1}}}

\newcommand{\Lagrangian}[1]{\mathcal{L}_{\text{#1}}}

\newcommand{\X}{\Tensor{X}}
\newcommand{\Y}{\Tensor{Y}}

\newcommand{\ProdMani}{\ThreeMani\times\mathbb{R}_t}

\newcommand{\JTens}{\Tensor{J}}

\newcommand{\FourVeloNullDust}{\Tensor{l}}

\newcommand{\NumDens}{n_{\BaseSpace}}

\newcommand{\Dx}{\dd x}
\newcommand{\Dy}{\dd y}
\newcommand{\Dv}{\dd v}
\newcommand{\Du}{\dd u}

\newcommand{\Flat}[1]{\prescript{\flat}{}{#1}}
\newcommand{\Twist}{\omega}
\newcommand{\Expansion}{\Theta}
\newcommand{\Shear}{\sigma}

\newcommand{\WeylScal}[1]{\Psi_{#1}}

\newcommand{\ComplNullVect}{\Tensor{m}}

\newcommand{\EinConst}{\kappa}

\newcommand{\TotNumStrings}{N}
\newcommand{\Genus}{k}

\newcommand{\ContConnect}{D}

\newcommand{\Kozaki}{%
  \author{Hiroshi \surname{Kozaki}}
  \email{kozaki@ishikawa-nct.ac.jp}
  \affiliation{Department of General Education, National Institute of Technology, Ishikawa College, Ishikawa 929-0392, Japan}
}

\newcommand{\Koike}{%
  \author{Tatsuhiko Koike} 
  \email{koike@phys.keio.ac.jp} 
  \affiliation{Department of Physics and Quantum Computing Center, Keio University, Yokohama 223-8522, Japan} 
  \affiliation{Research and Education Center for Natural Sciences, Keio University, Yokohama 223-8521, Japan} 
}

\newcommand{\Ishihara}{%
  \author{Hideki Ishihara}
  \email{h.ishihara@omu.ac.jp}
  \affiliation{Nambu Yoichiro Institute of Theoretical and Experimental Physics, Osaka Metropolitan University, Osaka 558-8585, Japan}
  \affiliation{Osaka Central Advanced Mathematical Institute, Osaka Metropolitan University, Osaka 558-8585, Japan}
}

\newcommand{\Morisawa}{%
  \author{Yoshiyuki Morisawa}
  \email{morisawa@omu.ac.jp}
  \affiliation{Osaka Central Advanced Mathematical Institute, Osaka Metropolitan University, Osaka 558-8585, Japan}
}

\begin{document}

\title{%
  Spacetime constructed from a contact manifold \\ with a degenerate metric
}

\Kozaki
\Ishihara
\Koike
\Morisawa

\preprint{OCU-PHYS-599}
\preprint{AP-GR-197}
\preprint{NITEP-220}

\begin{abstract}
   We construct a four-dimensional spacetime using a three-dimensional contact manifold equipped with a degenerate metric. 
   The degenerate metric is set to be compatible with the contact structure.
   The compatibility condition is defined in this paper.
   Our construction yields a Ricci tensor of a particularly simple form, 
   which leads to a solution of {\EinEq} with a null dust and cosmic strings.
   The solution includes two arbitrary functions: the energy density of the null dust and the number density of the cosmic
   strings. 
   When there exist the cosmic strings, the spacetime is of Petrov type D. 
   Otherwise, the spacetime is conformally flat. 
   For some simple matter densities, we examine {\EinEq} in detail.
\end{abstract}

\maketitle

\section{Introduction}

Exact solutions of {\EinEq} provide insight into gravitation and cosmology. 
The solutions are often constructed under the assumption that a spacetime has some high symmetry. 
The Schwarzschild solution is assumed to be static and spherically symmetric.
The Friedmann-Lemaître-Robertson-Walker universe is assumed to have spatial homogeneity and isotropy. 
Even if the assumption of high symmetry is relaxed, solutions are still obtained. 
An example is the Lemaître-Tolman-Bondi (LTB) solution, 
which describes dynamical and spherical symmetric spacetimes filled with a dust. 
A characteristic of the LTB solution is that it admits an arbitrary function,
which determines the energy density of the dust.

Contact manifolds or contact structures appear in various areas of physics. 
The most famous example is Hamiltonian mechanics \cite{arnol2013mathematical,GEIGES200125}. 
Others are thermodynamics \cite{MRUGALA1991109, GEIGES200125}, optics \cite{GEIGES200125}, and electromagnetism \cite{Dahl:2004}. 
In the study of magnetic trajectories in curved spaces, 
contact structures are effectively used \cite{Cabrerizo_2009, Inoguchi2019, Romaniuc2021}.
In the context of AdS/CFT correspondence,  
a special class of contact manifolds, called the Sasakian manifolds, play an important role \cite{Boyer:2007}. 
Recently, a contact structure has been found in the classical dynamics of a Nambu-Goto string \cite{Kozaki:2023vlu}.

The contact structures are also found to be useful to construct inhomogeneous solutions of {\EinEq}. 
Indeed, inhomogeneous generalizations of the Einstein static universe are obtained using three-dimensional contact manifolds
\cite{Ishihara:2021gty}. 
These solutions also have an arbitrary function, as in the case of the LTB solution.

A contact structure describes a twisting property of the system. 
In the classical dynamics of a Nambu-Goto string \cite{Kozaki:2023vlu}, the string worldsheet has a {\Nonvanishing} twist potential, 
which implies that the worldsheet is twisting along a direction.
The inhomogeneous generalizations of ESU \cite{Ishihara:2021gty} admit a twisting geodesic vector field, 
which is used to describe a fluid with vorticity. 
The twisting of the Gödel-type solutions is also understood in terms of the contact structure  \cite{Ishihara:2021abn}.

Another key to constructing exact solutions of {\EinEq}, apart from symmetry,
is a shear-free null geodesic congruence \cite{stephani_exact_2003,griffiths2009exact}. 
Solutions with such a congruence include {\Ppwaves}, Kundt solutions, and Robinson-Trautmann solutions. 
The {\Ppwaves}, which represent the plane-fronted gravitational waves with parallel rays, 
have also been attracting attention in string theories (e.g. \cite{Horowitz:1989bv,David_Berenstein_2002}). 
The shear-free null geodesic congruences of all the three solutions are twist-free. 
Solutions with twisting shear-free null geodesic congruences are also found \cite{Bini:2018gbq,Rosquist:2018ore}.

In this paper, we construct a four-dimensional spacetime from a three-dimensional contact manifold with a degenerate metric. 
For this aim, we first define a degenerate metric that is compatible with the contact structure. 
The metric compatibility has been considered for Riemannian as well as pseudo-Riemannian metrics
\cite{sasaki_differentiable_1960,takahashi_sasakian_1969,calvaruso_contact_2010, calvaruso_contact_2011}. 
We will extend this to degenerate metrics. 
Let $\ThreeMani$ be the contact manifold and $\ContMet$ be the compatible degenerate metric. 
We construct the spacetime manifold as $\R_u \times \ThreeMani$, where $u$ denotes a coordinate on $\R$.  
The spacetime metric is given by 
\begin{equation}
   \FourMet = - \ContForm \otimes \Du - \Du \otimes \ContForm + a^2(u) \ContMet,
\end{equation}
where $\ContForm$ is the $1$-form characterizing the contact structure and $a(u)$ is a function. 
This construction has three distinctive properties. 
The first is that the metric yields a Ricci tensor of a particularly simple form, 
which leads to a solution of {\EinEq} with a null dust and cosmic strings. 
This solution includes arbitrary functions, as in the case of the LTB solution. 
The second is that the spacetime admits two characteristic null vector fields: 
a covariantly constant null vector field and a null vector field generating a twisting shear-free null geodesic congruence. 
The existence of the former null vector field implies that our spacetime \Rev{falls into the Kundt class} 
\cite{stephani_exact_2003,griffiths2009exact}. 
The third is that the spacetime is of Petrov type D or conformally flat, which may be contributed by the null geodesic congruence. 
Using {\EinEq}, we find that the spacetime is of Petrov type D if and only if the cosmic strings exist; otherwise, it is conformally flat. 
These properties illustrate that our construction, which employs a contact manifold with a degenerate metric, is intriguing.

The paper is organized as follows. 
In the next section, we first outline the contact manifolds and the Riemannian and pseudo-Riemannian metrics compatible with the
contact structure.  
Then, we extend the notion of the metric compatibility to degenerate metrics. 
In Sec.~\ref{sec:ConstructionOfSpacetime}, 
we construct a spacetime using a three-dimensional contact manifold with a compatible degenerate metric. 
The geometrical properties are clarified. 
In Secs.~\ref{sec:EinsteinEquation},\ref{sec:GeometryOfBaseSpace}, and \ref{sec:EvolutionOFWarpFactor}, 
we examine {\EinEq}.
Sec.~\ref{sec:Conclusion} is devoted to conclusions.

\section{%
  \label{sec:DegenerateMetricOnContactManifold}%
  Degenerate metric on a contact manifold
}

\subsection{%
  \label{subsec:Preliminaries}%
  Preliminaries%
}

\subsubsection{%
  \label{subsubsec:ContactManifold}%
  Contact manifold%
}

A contact manifold $\ContMani$ is a $(2n+1)$-dimensional manifold with a $1$-form $\ContForm$ which satisfies the condition
\cite{Boothby:1958}: 
\begin{equation}
   \label{eq:DefContact}
   \ContForm \wedge \underbrace{\dContForm \wedge \dots \wedge \dContForm}_{\text{$n$ factors}} \neq 0
   \quad
   \text{at all points in $\ContMani$}. 
\end{equation}
This $1$-form is called the contact form \cite{Blair:1976}.
The condition \eqref{eq:DefContact} implies that, at each point $p \in \ContMani$, 
the skew-symmetric bilinear form $\dContForm_p$ is {\Nondegenerate} on the subspace $\ker \ContForm_p \subset T_p\ContMani$ 
(see Eq.~\eqref{eq:NondegenExtDContactForm}). 
Thus, $\ker \ContForm_p$ is a symplectic vector space of dimension $2n$.
However, the subbundle $\bigcup_{p\in\ContMani} \ker \ContForm_p$ is not integrable,
i.e., there is no submanifold such that, at each point $p$ on it, the tangent space coincides with $\ker\ContForm_p$,
because it follows from the condition \eqref{eq:DefContact} that $\ContForm \wedge \dContForm \neq 0$. 
The condition \eqref{eq:DefContact} is an assumption of Darboux's theorem \cite{Blair:1976}. 
Therefore, the contact manifold $\ContMani$ admits local coordinates $v$, $x^1,\dots,x^n$, $y^1, \dots, y^n$ such that 
\begin{equation}
  \ContForm = \Dv + \sum_{i = 1}^n x^i \Dy^i. 
\end{equation}
We will call these coordinates the \Emph{Darboux coordinates} in this paper.

 In a contact manifold $\ContMani$ with a contact form $\ContForm$,
there exists a unique vector field $\ReebVect$ such that \cite{Blair:1976}
\begin{subequations}
   \begin{align}
     \label{eq:DefReeb}
     \IntProd_{\ReebVect} \dContForm &\coloneq \dContForm(\ReebVect, \EmptySlot) = 0,  
     \\
     \label{eq:NormalizationReeb}
     \IntProd_{\ReebVect} \ContForm  &\coloneq \ContForm(\ReebVect)              = 1. 
   \end{align}
\end{subequations}
This vector field is called the Reeb vector field. 
In the Darboux coordinates, the Reeb vector field is simply written as 
\begin{equation}
  \label{eq:ReebVectorInDarbouCoordinates}
  \ReebVect = \pdv{v}. 
\end{equation}
For the contact form $\ContForm$ and the Reeb vector field $\ReebVect$, 
a $(1,1)$ tensor field $\PhiTens$ is defined so that \cite{sasaki_differentiable_1960}
\begin{equation}
   \label{eq:DefPhiTensor}
   \PhiTens^2 = - \Tensor{1} + \ReebVect \otimes \ContForm. 
\end{equation}
This tensor field satisfies the following equations \cite{Blair:1976} 
\begin{subequations}
   \begin{align}
     \label{eq:PhiXi=0}
     \PhiTens \ReebVect       &\coloneq \PhiTens(\EmptySlot , \ReebVect) = 0,
     \\
     \label{eq:EtaPhi=0}
     \ContForm \circ \PhiTens &\coloneq \PhiTens(\ContForm, \EmptySlot) = 0.
   \end{align}
\end{subequations}
Equations~\eqref{eq:DefPhiTensor} and \eqref{eq:EtaPhi=0} shows that, at each point $p \in \ContMani$,
the linear map $\PhiTens_p: T_p \ContMani \to T_p \ContMani$ becomes an almost complex structure in $\ker \ContForm_p$, 
which is a symplectic vector space with the symplectic form $\dContForm_p$.

\subsubsection{%
  \label{subsubsec:NondegenerateMetricOnContactManifold}%
  Nondegenerate metric on a contact manifold
}

A {\Nondegenerate} metric $\ContMet$ on a contact manifold $\ContMani$ is said to be compatible with the contact structure
if the following equations are satisfied
\cite{sasaki_differentiable_1960,takahashi_sasakian_1969,calvaruso_contact_2010, calvaruso_contact_2011}: 
\begin{subequations}
   \begin{align}
     \label{eq:DefCompatMetric}
     \ContMet(\PhiTens \X, \PhiTens \Y) &= \ContMet(\X, \Y) - \varepsilon \, \ContForm(\X) \ContForm(\Y),
     \\
     \label{eq:DefContactMetric}
     \ContMet(\X, \PhiTens \Y)          &= \dContForm (\X, \Y), 
   \end{align}
\end{subequations}
where $\varepsilon = \pm 1$, and $\X$ and $\Y$ are arbitrary vector fields.
In terms of the symplectic vector space $\ker \ContForm_p$ at each point $p \in \ContMani$, 
these equations imply that the almost complex structure $\PhiTens_p$ is compatible with the symplectic structure $\dContForm_p$
if the restriction of $\ContMet_p$ on $\ker\ContForm_p$ is positive definite. 
We will call the metric satisfying Eqs.~\eqref{eq:DefCompatMetric} and \eqref{eq:DefContactMetric} the \Emph{contact metric}.
The contact metric links the Reeb vector field $\ReebVect$ to the contact form $\ContForm$ as follows: 
\begin{equation}
   \label{eq:MetDualContactFormReeb}
   \ContMet(\ReebVect, \EmptySlot) = \varepsilon \ContForm. 
\end{equation}
This leads to the normalization of the Reeb vector field $\ContMet(\ReebVect, \ReebVect) = \varepsilon$. 
A remarkable feature of the contact metric is that the Reeb vector field $\ReebVect$ is a geodesic vector field \cite{Blair:1976}.

When the Reeb vector field $\ReebVect$ is a Killing vector field,
i.e., $\LieDeriv_{\ReebVect} \ContMet = 0$ where $\LieDeriv_{\ReebVect}$ denotes the Lie derivative along $\ReebVect$,
the contact manifold is called the K-contact manifold \cite{sasaki_differentiable_1961,hatakeyama_properties_1963}.

\subsection{%
  \label{subsec:Extension}%
  Extension to degenerate metrics%
}

We extend the notion of the metric compatibility to degenerate metrics. 

Let $\ContMet$ be a degenerate metric on a contact manifold $\ContMani$. 
If the degenerate metric satisfies Eqs.~\eqref{eq:DefCompatMetric} and \eqref{eq:DefContactMetric} with $\varepsilon = 0$,
we define $\ContMet$ to be compatible with the contact structure, and call $\ContMet$ the \Emph{contact metric} as in the case of {\Nondegenerate} metrics.  
This extension does not alter the essence of the compatibility. 
That is, in the symplectic vector space $\ker \ContForm_p$ at each point $p \in \ContMani$,
the almost complex structure $\PhiTens_p$ is compatible with the symplectic structure $\dContForm_p$
if the restriction of $\ContMet_p$ on $\ker \ContForm_p$ is positive definite. 
Furthermore, this extension leads to the same feature that the Reeb vector field $\ReebVect$ is a geodesic vector field
(see Appendix~\ref{sec:ProofReebIsGeodesic} for a proof). 
A difference arises in the relation between the metric and the Reeb vector field. 
In fact, from Eq.~\eqref{eq:DefCompatMetric} with $\varepsilon = 0$, we have 
\begin{equation}
   \label{eq:DegenContactMetric}
   \ContMet(\ReebVect, \EmptySlot) = 0, 
\end{equation}
whereas, when $\ContMet$ is {\Nondegenerate}, $\ContMet(\ReebVect,\EmptySlot)$ gives the contact form $\ContForm$ as shown
in Eq.~\eqref{eq:MetDualContactFormReeb}.  
Equation~\eqref{eq:DegenContactMetric} shows that the Reeb vector field gives a degenerate direction of the metric. 
It will be shown in the next subsection that this is the only degenerate direction.

Even when a contact metric is degenerate,
we will call the contact manifold the K-contact manifold if the Reeb vector field is a Killing vector field.

\subsection{%
  \label{subsec:ConstructionContactMetric}
  Construction of a contact metric 
}

We outline a process of constructing a contact metric from a given contact form $\ContForm$ for all the cases $\varepsilon = -1, 0, 1$. 
We then present an actual construction in three dimensions.

Let $\ReebVect$ be the Reeb vector field of the contact form $\ContForm$. 
We define a local frame $\qty{\ReebVect, \FrmFld_1, \dots, \FrmFld_{2n}}$
so that $\FrmFld_1, \dots, \FrmFld_{2n}$ span $\ker \ContForm_p$ at each point $p \in \ContMani$, namely 
\begin{equation}
   \ContForm (\FrmFld_I) = 0 \quad (I = 1,\dots,2n). 
\end{equation}
As a corresponding coframe, we take $\qty{\ContForm, \CofFld^1, \dots, \CofFld^{2n}}$ in which  
\begin{equation}
   \CofFld^I(\ReebVect) = 0, 
   \qquad
   \CofFld^I(\FrmFld_J) = \delta^I {}_J
   \quad
   (I,J = 1,\dots, 2n). 
\end{equation}
With respect to these frame and coframe, the exterior derivative of $\ContForm$, which satisfies Eq.~\eqref{eq:DefReeb}, is expressed as 
\begin{equation}
   \label{eq:ExpdContForm}
   \dContForm = \dContFormComp_{IJ} \, \CofFld^I \otimes \CofFld^J,
   \qquad
   \dContFormComp_{IJ} \coloneq \dContForm(\FrmFld_I, \FrmFld_J).
\end{equation}
The $(1,1)$ tensor field $\PhiTens$, which is defined to satisfy Eq.~\eqref{eq:DefPhiTensor}, is written as follows:
\begin{equation}
   \PhiTens = \PhiTensComp^I{}_J \, \FrmFld_I \otimes \CofFld^J,
   \qquad
   \PhiTensComp^I{}_J \coloneq \PhiTens(\CofFld^I,\FrmFld_J), 
\end{equation}
where we have used Eqs.~\eqref{eq:PhiXi=0} and \eqref{eq:EtaPhi=0}.
It follows from Eq.~\eqref{eq:DefPhiTensor} that $\PhiTensComp^I{}_J$ must satisfy 
\begin{equation}
   \label{eq:ComplexityPhiTensor}
   \PhiTensComp^I{}_J \, \PhiTensComp^J{}_K  = - \delta^I{}_K. 
\end{equation}
The contact metric $\ContMet$ is expanded as
\begin{equation}
   \label{eq:ExpContMet}
   \ContMet = \varepsilon \ContForm \otimes \ContForm + \ContMetComp_{IJ} \, \CofFld^I \otimes \CofFld^J, 
\end{equation}
where we have used Eq.~\eqref{eq:MetDualContactFormReeb} or Eq.~\eqref{eq:DegenContactMetric}. 
It follows from the compatible condition, namely Eqs.~\eqref{eq:DefCompatMetric} and \eqref{eq:DefContactMetric}, that
$\ContMetComp_{IJ}$ must satisfy the following equations:
\begin{subequations}
   \begin{align}
     \label{eq:CondCompatMetricComp}
     \ContMetComp_{IJ} \PhiTensComp^I{}_K \PhiTensComp^J{}_L &= \ContMetComp_{KL},
     \\
     \label{eq:CondContactMetricComp}
     \ContMetComp_{IJ} \PhiTensComp^J{}_K                    &= \dContFormComp_{IK}.
   \end{align}
\end{subequations}
From Eqs.~\eqref{eq:ComplexityPhiTensor} and \eqref{eq:CondContactMetricComp},
$\ContMetComp_{IJ}$ is obtained as 
\begin{equation}
   \label{eq:ContMetFromPhiTens}
   \ContMetComp_{IJ}   = - \dContFormComp_{IK}\PhiTensComp^K{}_J.
\end{equation}
We note that the right hand side is not necessarily symmetric about the indices $I$ and $J$.
Therefore, we have to take $\PhiTensComp^I{}_J$ so that $\ContMetComp_{IJ}$ given by Eq.~\eqref{eq:ContMetFromPhiTens} is
symmetric.

In three dimensions, i.e., $2n + 1 = 3$, Eq.~\eqref{eq:ContMetFromPhiTens} always gives a symmetric $\ContMetComp_{IJ}$,
and hence the contact metric is easily obtained. 
Indeed, in three dimensions, $\dContFormComp_{IJ}$ and $\PhiTensComp^I{}_J$ are generally given by 
\begin{equation}
   \dContFormComp_{IJ} = \mqty[0 & \alpha \\ -\alpha & 0],
   \qquad
   \PhiTensComp^I{}_J = \mqty[f  & c \\ -b & -f], 
\end{equation}
where $\alpha$ is a {\Nonvanishing} function, and $b$, $c$ , and $f$ are functions satisfying 
\begin{equation}
   \label{eq:CondPhiTensorFunc}
   f^2 - bc = -1. 
\end{equation} 
As a result, Eq.~\eqref{eq:ContMetFromPhiTens} certainly yields a symmetric $\ContMetComp_{IJ}$: 
\begin{equation}
   \label{eq:ContactMetricComp}
   \ContMetComp_{IJ} = \alpha \mqty[ b & f \\ f & c ].
\end{equation}
Using this $\ContMetComp_{IJ}$ in Eq.~\eqref{eq:ExpContMet}, we obtain a contact metric. 
The definiteness of $\ContMetComp_{IJ}$ is determined by the functions $\alpha$ and $b$.  
It follows from Eq.~\eqref{eq:CondPhiTensorFunc}
that $\ContMetComp_{IJ}$ is positive definite if and only if $\alpha b > 0$.

It is shown that, at each point $p \in \ContMani$, the bilinear form $\dContForm_p$ is 
{\Nondegenerate} on the subspace $\ker \ContForm_p$ by using Eq.~\eqref{eq:ExpdContForm}. 
Indeed, applying the $(2n+1)$-form $\ContForm\wedge\dContForm\wedge\dots\wedge\dContForm$ to the frame fields
$\ReebVect, \FrmFld_1, \dots \, , \FrmFld_{2n}$, 
we have
\begin{equation}
   \label{eq:NondegenExtDContactForm}
   \det \dContFormComp_{IJ} \neq 0. 
\end{equation}
It is also shown that the Reeb vector field $\ReebVect$ gives the only degenerate direction of $\ContMet$ with $\varepsilon = 0$. 
Indeed, it follows from Eqs.~\eqref{eq:ComplexityPhiTensor}, \eqref{eq:CondContactMetricComp},
and \eqref{eq:NondegenExtDContactForm} that
\begin{equation}
  \det \ContMetComp_{IJ} = \pm \det \dContFormComp_{IJ} \neq 0. 
\end{equation}
This means that $\ContMetComp_{IJ}$ is invertible. 
Consequently, if there exists a vector field $\X$ satisfying $\ContMet(\X, \EmptySlot) = 0$, 
we can show that $\X$ is a scalar multiple of $\ReebVect$.

\subsection{%
  \label{subsec:ContactMetricThreeDimensions}%
  Three dimensional K-contact manifold}

We explore the geometric properties of a three-dimensional K-contact manifold $\ThreeMani$ with
$\varepsilon = -1, 0, 1$, 
where the Reeb vector field $\ReebVect$ is assumed to be a Killing vector field.

First, we write down the contact metric in the Darboux coordinates $v, x, y$,
with which the contact form and the Reeb vector field is written as follows: 
\begin{equation}
   \label{eq:FrameCoframeDarbouxCoord1}
  \ContForm = \Dv + x \Dy, \qquad \ReebVect = \pdv{v}. 
\end{equation}
In the Darboux coordinates, we take the following frame fields and coframe fields: 
\begin{equation}\label{eq:FrameCoframeDarbouxCoord2}
   \FrmFld_1 = \pdv{x}, 
   \qquad
   \FrmFld_2 = \pdv{y} - x \pdv{v},
   \qquad
   \CofFld^1 = \Dx,
   \qquad
   \CofFld^2 = \Dy. 
\end{equation}
Then, we have 
\begin{align}
  \label{eq:dContFormIJThreeDim}
  \dContFormComp_{IJ} = \dContForm(\FrmFld_{I}, \FrmFld_{J}) = \mqty[0  & 1 \\ -1 & 0], 
\end{align}
and hence $\ContMetComp_{IJ}$ is obtained from Eq.~\eqref{eq:ContactMetricComp} as 
\begin{equation}
   \ContMetComp_{IJ} = \mqty[b & f \\ f & c]. 
\end{equation}
The functions $b$, $c$, and $f$ do not depend on $v$
because the Reeb vector field $\ReebVect = \pdv*{v}$ is assumed to be a Killing vector field. 
In this paper, we will assume that $b > 0$ so that $\ContMetComp_{IJ}$ is positive definite, 
and we express $b$ as $e^{2 \Omega}$ where  $\Omega$ is an arbitrary function of $x,y$. 
In this setting, Eq.~\eqref{eq:CondPhiTensorFunc} yields $c = e^{-2 \Omega} (1 + f^2)$. 
Then, from Eq.~\eqref{eq:ExpContMet}, the contact metric $\ContMet$ is explicitly written as follows: 
\begin{equation}
   \label{eq:ContactMetricDarbouxCoord}
   \ContMet = \varepsilon (\Dv + x \Dy)^2 + e^{2 \Omega} \, \Dx^2 + 2 f \Dx \Dy + e^{-2\Omega}(1 + f^2) \, \Dy^2.
\end{equation}
We remark that $\ThreeMani$ has a natural fiber bundle structure $\ThreeMani \to \BaseSpace$,
where $\BaseSpace$ is the quotient of $\ThreeMani$ by the isometry group generated by the Reeb vector $\ReebVect$. 
The metric on the base space $\BaseSpace$ is given by 
\begin{equation}
   \label{eq:BaseMet}
   \BaseMet = e^{2 \Omega} \, \Dx^2 + 2 f \Dx \Dy + e^{-2 \Omega}(1 + f^2) \, \Dy^2. 
\end{equation}

Next, we consider the Ricci tensor for the cases $\varepsilon = \pm 1$. 
The Ricci tensor of the metric~\eqref{eq:ContactMetricDarbouxCoord} is calculated as follows: 
\begin{equation}
   \label{eq:RicciTensorOnK-contact}
   \ThreeRicTens
   =
   \frac{\BaseRicScal - \varepsilon}{2} \ContMet
   +
   \varepsilon \frac{2 \varepsilon - \BaseRicScal}{2} \ContForm \otimes \ContForm, 
\end{equation}
where $\BaseRicScal$ is the Ricci scalar of the base space $\BaseSpace$ given by 
\begin{equation}
  \label{eq:BaseRicSca}
  \BaseRicScal 
  =
  -
  \pdv[2]{x} \qty[(1 + f^2) e^{-2 \Omega}]
  -
  \pdv[2]{y} e^{2 \Omega}
  +
  2\qty(\pdv{f}{x} \pdv{\Omega}{y} - \pdv{f}{y} \pdv{\Omega}{x} + \pdv{f}{x}{y}).
\end{equation}
Equation~\eqref{eq:RicciTensorOnK-contact} implies that any three-dimensional K-contact manifold is $\eta$-Einstein
\cite{okumura_remarks_1962}.%
\footnote{Here, we call a contact manifold $\eta$-Einstein when the Ricci tensor is given as the sum of the contact metric
  $\ContMet$ and $\ContForm \otimes \ContForm$ each multiplied by a function. 
  In higher dimensional K-contact manifolds, the coefficient functions are shown to be constant.}
The contravariant Ricci tensor $\ThreeContraRicTens$, 
whose indices are raised by the inverse metric, also takes a characteristic form: 
\begin{equation}
   \label{eq:ContraRicciTensorOnK-contact}
   \ThreeContraRicTens
   =
   \frac{\BaseRicScal - \varepsilon}{2} \ContMetComp^{IJ} \FrmFld_{I} \otimes \FrmFld_J
   +
   \frac{1}{2} \ReebVect \otimes \ReebVect, 
\end{equation}
where $\ContMetComp^{IJ}$ is the inverse of $\ContMetComp_{IJ}$. 
We note that, if the metric is degenerate, i.e., $\varepsilon = 0$, 
the Levi-Civita connection and the Ricci tensor are not uniquely determined by the metric.

We remark that we can define a Sasakian structure in a contact manifold with a degenerate metric
(see Appendix~\ref{sec:SasakianStructure} for detailed discussion). 
It can be shown that, in a three dimensional contact manifold with a degenerate metric, 
being K-contact is equivalent to being Sasakian.

\section{%
  \label{sec:ConstructionOfSpacetime}
  Construction of a spacetime 
}

\begin{figure}[h]
   \centering
   \includegraphics[]{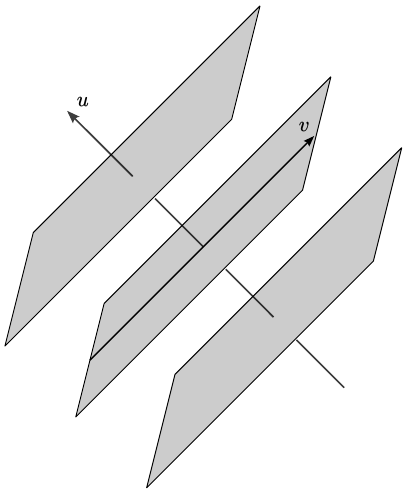}
   \caption{
     A rough sketch of the spacetime construction.
     The null hypersurfaces, $u = \text{const.}$,
     which are homothetic to a three-dimensional K-contact manifold $\ThreeMani$ with a degenerate metric,
     are homothetically stacked along the null direction $u$.} 
   \label{fig:Configuration}
\end{figure}

We construct a four-dimensional spacetime using a three-dimensional degenerate contact manifold. 
Let $\ThreeMani$ be a three-dimensional K-contact manifold with a contact form $\ContForm$ and a degenerate contact metric $\ContMet$. 
The spacetime manifold is topologically a product of the real line $\R$ and $\ThreeMani$.
Let $u$ be a coordinate on $\R$. 
We define a spacetime metric as 
\begin{equation}
   \label{eq:MetricConstruction}
   \FourMet = - \Du \otimes \ContForm - \ContForm \otimes \Du + a^2 \ContMet, 
\end{equation}
where $a$ is a function of $u$ which we call the warp factor. 
Using the Darboux coordinates $v,x,y$ on $\ThreeMani$, we can write the metric as 
\begin{equation}
   \label{eq:FourMetricComp}
   \FourMet = -2 \Du (\Dv + x \Dy) + a^2 \qty[e^{2\Omega} \, \Dx^2 + 2 f \Dx \Dy + e^{-2\Omega}(1 + f^2) \, \Dy^2], 
\end{equation}
where $\Omega$ and $f$ are functions of $x$ and $y$.
We see that the coordinate vector field along $\R$, i.e., 
\begin{equation} 
   \pdv{u} \eqcolon \NullFrmFldRL, 
\end{equation}
is null and that the hypersurfaces, $u = \text{const.}$, are also null and homothetic to $\ThreeMani$.  
Thus, the spacetime is constructed by homothetically stacking $\ThreeMani$ along the null direction as shown in 
Fig.~\ref{fig:Configuration}.  
The other null coordinate vector field, 
\begin{equation}
   \pdv{v} \eqcolon \NullFrmFldReeb, 
\end{equation}
is a Killing vector field. 
It should be noted that the metric dual $1$-form of $\NullFrmFldReeb$, 
which will be denoted by $\Flat{\NullFrmFldReeb}$, 
is not related to $\ContForm (= \Dv + x \Dy)$. 
In fact, it follows that 
\begin{equation}
   \Flat{\NullFrmFldReeb} \coloneq \FourMet(\NullFrmFldReeb, \EmptySlot) = - \dd  u.    
\end{equation}
This may confuse the readers who are familiar with the contact manifolds with a {\Nondegenerate} metric
(cf. Eq.~\eqref{eq:MetDualContactFormReeb}).
The metric dual of $\NullFrmFldRL$ gives the contact form $\ContForm$; 
indeed, it follows that 
\begin{equation}
   \Flat{\NullFrmFldRL} \coloneq \FourMet(\NullFrmFldRL, \EmptySlot) = - \ContForm. 
\end{equation}
As used above, the flat symbol $\flat$ will be used for vectors and contravariant tensors 
when all the indices are lowered by the metric. 
On the other hand, as previously used in Eq.~\eqref{eq:ContraRicciTensorOnK-contact}, 
the sharp symbol $\sharp$ will be used for $1$-forms and covariant tensors when all the indices are raised by the inverse metric.

A striking feature of the metric \eqref{eq:MetricConstruction} is that the Ricci tensor takes the following simple form: 
\begin{equation}
  \label{eq:FourDimensionalRicciTensor}
  \FourRicTens = \frac{\BaseRicScal}{2} \ContMet + \qty(\frac{1}{2a^4} -2 \frac{a''}{a}) \Du \otimes \Du,
\end{equation}
where $\BaseRicScal$ is the Ricci scalar of the base space $\BaseSpace$ given by Eq.~\eqref{eq:BaseRicSca},
and the prime denotes the differentiation with $u$. 
The contravariant version of the Ricci tensor  $\FourContraRicTens$ is written as 
\begin{equation}
   \label{eq:ContravariantFourRicciTensor}
   \FourContraRicTens
   =
   \frac{\BaseRicScal}{2a^4} \ContMetComp^{IJ} \FrmFld_I \otimes \FrmFld_J 
   +
   \qty(\frac{1}{2a^4} -2 \frac{a''}{a}) \NullFrmFldReeb \otimes \NullFrmFldReeb. 
\end{equation}
In the simple case where $a(u) = 1$, 
the right hand side coincides with the expression of the three-dimensional Ricci tensor with $\varepsilon = 0$ in 
Eq.~\eqref{eq:ContraRicciTensorOnK-contact}.
In three-dimensions, a {\Nondegenerate} K-contact manifold is always $\eta$-Einstein, 
where the contravariant Ricci tensor always takes the form of Eq.~\eqref{eq:ContraRicciTensorOnK-contact}. 
The corresponding property of a degenerate K-contact manifold emerges when the manifold is realized as a null hypersurface in a
four-dimensional spacetime constructed by Eq.~\eqref{eq:MetricConstruction}.

Another important feature of the metric \eqref{eq:MetricConstruction} concerns the two null vector fields $\NullFrmFldReeb$
and $\NullFrmFldRL$. 
The null vector field $\NullFrmFldReeb$ is covariantly constant,  $\nabla \NullFrmFldReeb = 0$, 
where $\nabla$ denotes the Levi-Civita connection. 
This implies that \Rev{$\NullFrmFldReeb$ generates null geodesic congruence with vanishing optical scalars,
  and hence our spacetime is in the Kundt class  \cite{stephani_exact_2003,griffiths2009exact}.}
On the other hand, the null vector field $\NullFrmFldRL$ generates twisting shear-free null geodesic congruence. 
It is easily checked that  $\nabla_{\NullFrmFldRL} \NullFrmFldRL = 0$ 
and that the expansion $\Expansion$, the shear $\Shear$, and the twist $\Twist$ are obtained as 
\begin{equation}
   \Expansion = \frac{a'}{a}, \qquad \Shear \bar{\Shear} = 0, \qquad \Twist^2 = \frac{1}{4a^4}. 
\end{equation}

Our spacetime is of Petrov type D or conformally flat.
This can be seen by calculating the Weyl scalars $\WeylScal{0}, \dots, \WeylScal{4}$
\cite{stephani_exact_2003,griffiths2009exact} using a null complex tetrad 
$(\ComplNullVect, \bar{\ComplNullVect}, \NullFrmFldRL, \NullFrmFldReeb)$, 
where the complex null vector field $\ComplNullVect$ is defined by 
\begin{equation}
   \ComplNullVect \coloneq \frac{i e^{\Omega}}{\sqrt{2}a} \qty[x \pdv{v} + (f - i) e^{-2\Omega} \pdv{x} - \pdv{y}]. 
\end{equation}
We find that the only {\Nonvanishing} Weyl scalar is
\begin{equation}
   \label{eq:NonvanishingWeyl}
   \
   \WeylScal{2} = -\frac{\BaseRicScal}{12 a^2}. 
\end{equation}
\Rev{If $\BaseRicScal \neq 0$, the spacetime is of Petrov Type D} (see Table~4.2 of \cite{stephani_exact_2003}).
\Rev{The two repeated principal null directions} 
are given by $\NullFrmFldRL$ and $\NullFrmFldReeb$. 
If $\BaseRicScal = 0$, all of the Weyl scalars vanish, and hence,
the spacetime is conformally flat (Petrov type O). 
\Rev{Furthermore, our spacetime is in the class of spacetimes with vanishing curvature invariants, 
  i.e., all scalars obtained by contracting a polynomial of the Riemann tensor and its covariant derivatives of any
  order vanish. This is seen by the form of the Ricci tensor \eqref{eq:FourDimensionalRicciTensor} and a theorem in
  \cite{Pravda:2002us}.}

\section{%
  \label{sec:EinsteinEquation}%
  {\EinEq}
} 
We will show that our spacetime gives a solution to {\EinEq} with a null dust and cosmic strings. 
We will write {\EinEq} in the following contravariant form: 
\begin{equation}
   \label{eq:EinsteinEquation}
   \EinTens = \EinConst \EneMomTens{}, 
\end{equation}
where $\EinTens$ is the contravariant Einstein tensor,
$\EneMomTens{}$ is the contravariant energy-momentum tensor, 
and $\EinConst$ is the Einstein constant. 
The key is that the contravariant Ricci tensor~\eqref{eq:ContravariantFourRicciTensor} gives $\EinTens$ the following form:  
\begin{equation}
   \label{eq:EinsteinTensor}
   \EinTens
   =
   \qty(\frac{1}{2a^4} - 2 \frac{a''}{a}) \NullFrmFldReeb \otimes \NullFrmFldReeb 
   +
   \frac{\BaseRicScal}{2a^2} \qty(\NullFrmFldRL \otimes \NullFrmFldReeb + \NullFrmFldReeb \otimes \NullFrmFldRL). 
\end{equation}
In the following Subsecs.~\ref{subsec:NullDust} and \ref{subsec:CosmicStrings}, 
we will show that the first term of $\EinTens$ is the contribution due to a null dust and the second term is that due to
cosmic strings.
In Subsec.~\ref{subsec:ReductionToSeparatedSystem}, 
we will write down {\EinEq} and find that the roles of the null dust and the cosmic strings are completely separated.

\subsection{%
  \label{subsec:NullDust}%
  Null dust
}

Null dust is a fluid that consists of massless particles moving in a null direction. 
Let the vector field $\FourVeloNullDust$ denote the null direction. 
Then, the contravariant energy-momentum tensor is given by \cite{stephani_exact_2003} 
\begin{equation}
   \label{eq:EnergyMomentumNullDust}
   \EneMomTens{\LablNullDust} = \EneDensNullDust \FourVeloNullDust \otimes \FourVeloNullDust, 
\end{equation}
where $\EneDensNullDust$ is the energy density of the null dust. 
It is easily checked that the covariant divergence $\nabla \cdot \EneMomTens{\LablNullDust}$ 
vanishes if and only if 
\begin{equation}
   \label{eq:ConditionForEOMNullDust}
   \nabla_{\FourVeloNullDust} \FourVeloNullDust
   =
   - \frac{%
     \nabla \cdot \qty(\EneDensNullDust \FourVeloNullDust)
   }{%
     \EneDensNullDust}
   \FourVeloNullDust. 
\end{equation}
This is the equation of motion for the null dust. 
This implies that the null vector field $\FourVeloNullDust$ must be geodesic.

In our spacetime, we consider a null dust moving along the null vector field $\NullFrmFldReeb$. 
Then, the energy-momentum tensor,
which is given by Eq.~\eqref{eq:EnergyMomentumNullDust} with $\FourVeloNullDust = \NullFrmFldReeb$, 
is compatible with the first term of the Einstein tensor \eqref{eq:EinsteinTensor}. 
The equation of motion \eqref{eq:ConditionForEOMNullDust} reads
\begin{equation}
   \label{eq:ConditionForEnergyDensityOfNullDust}
   0 = \NullFrmFldReeb(\EneDensNullDust) = \pdv{\EneDensNullDust}{v}, 
\end{equation}
where we have used that $\NullFrmFldReeb$ is covariantly constant, i.e.,  $\nabla\NullFrmFldReeb = 0$.

\subsection{%
  \label{subsec:CosmicStrings}
  Cosmic strings
}

We begin by considering the gravitational field coupled with a single cosmic string.
The action is given by
\begin{equation}
   \Action{} = \Action{\LablEinHil} + \Action{\LablCosmicString}, 
\end{equation}
where $\Action{\LablEinHil}$ is the Einstein-Hilbert action and 
$\Action{\LablCosmicString}$ is the action of the cosmic string that we assume to be the Nambu-Goto action. 
Let $\zeta^a~(a = 1,2)$ and $x^\mu~(\mu=0,\dots,3)$ be coordinates on the string worldsheet and the spacetime manifold. 
Let $x^\mu = X^\mu(\zeta^a)$ be the embedding of the string worldsheet. 
Then the string action $\Action{\LablCosmicString}$ is given by
\begin{align}
  \Action{\LablCosmicString}[g, X]
  &= 
    -\mu \int \sqrt{-\gamma} \, \dd[2]\zeta
    =
    \int \Lagrangian{\LablCosmicString}\sqrt{-g}\, \dd[4]x,
  \\
  \Lagrangian{\LablCosmicString}
  &\coloneq
    -\frac{\mu}{\sqrt{-g}}
    \int 
    \delta^4(x - X(\zeta))
    \sqrt{-\gamma}\,
    \dd[2]\zeta,
\end{align}
where $\mu$ is the tension of the cosmic string and $\gamma$ is the determinant of the worldsheet metric  
\begin{equation}
   \gamma_{ab} \coloneq g_{\mu\nu} \pdv{X^\mu}{\zeta^a}\pdv{X^\nu}{\zeta^b}. 
\end{equation}
The energy-momentum tensor of the cosmic string is given by 
\begin{equation}
   \label{eq:GeneralEnergyMomentumTensorString}
   T^{\mu\nu}_{\text{\LablCosmicString}}
   =
   \frac{2}{\sqrt{-g}} \fdv{\Action{\LablCosmicString}}{g_{\mu\nu}}
   =
   -\frac{\mu}{\sqrt{-g}}
   \int
   \delta^4(x - X(\zeta))
   \sqrt{-\gamma} \, \gamma^{ab} \pdv{X^\mu}{\zeta^a} \pdv{X^\nu}{\zeta^b}
   \dd[2]\zeta. 
\end{equation}

In our spacetime, where the metric is given by Eq.~\eqref{eq:FourMetricComp},  
we take the following worldsheet embedding: 
\begin{equation}\label{eq:Worldsheet}
   u(\zeta^a) = \zeta^1,
   \qquad
   v(\zeta^a) = \zeta^2,
   \qquad
   x(\zeta^a) = x_1, 
   \qquad
   y(\zeta^a) = y_1, 
\end{equation}
where $x_1$ and $y_1$ are constants. 
Then, from Eq.~\eqref{eq:GeneralEnergyMomentumTensorString}, the energy-momentum tensor is calculated as follows: 
\begin{equation}
   \label{eq:EnergyMomentumSingleString}
   \EneMomTens{\LablCosmicString}
   =
   \frac{\mu}{a^2}
   \delta(x - x_1) \delta(y - y_1)
   \qty(
   \NullFrmFldRL \otimes \NullFrmFldReeb
   +
   \NullFrmFldReeb \otimes \NullFrmFldRL
   )
   \eqcolon
   \EneMomTens[(x_1,y_1)]{\LablCosmicString}. 
\end{equation}
We note that the worldsheet embedding of Eq.~\eqref{eq:Worldsheet} satisfies the Nambu-Goto equation of motion. 
In particular, it is an exact solution of the Nambu-Goto strings with a null symmetry \cite{Kozaki:2023vlu}.

We extend the energy-momentum tensor \eqref{eq:EnergyMomentumSingleString} to multiple cosmic strings under the assumption that there is no interaction between the cosmic strings. 
Suppose that there are cosmic strings whose worldsheets are specified by $(x,y) = (x_1,y_1), \dots, (x_N, y_N)$.
Then the total energy-momentum tensor may be given by 
\begin{equation}
  \EneMomTens{\LablCosmicStrings} = \sum_{i=1}^N \EneMomTens[(x_i,y_i)]{\LablCosmicString}. 
\end{equation}
The cosmic strings are identified with points on $\BaseSpace$.
Let $\NumDens(x,y)$ be a number density function of the cosmic strings on $\BaseSpace$. 
Then the total energy-momentum tensor may be given by 
\begin{align}
  \label{eq:EnergyMomentumCosmicStrings}
  \EneMomTens{\LablCosmicStrings}
  &=
    \iint_{\BaseSpace}
    \NumDens(x',y')
    \EneMomTens[(x',y')]{\LablCosmicString}
    \sqrt{\det \BaseMet} \, \Dx' \Dy'
    \notag
  \\
  &=
   \frac{\mu \NumDens(x,y)}{a^2}
   \qty(
   \NullFrmFldRL \otimes \NullFrmFldReeb
   +
   \NullFrmFldReeb \otimes \NullFrmFldRL
   ). 
\end{align}
This is clearly compatible with the second term of the Einstein tensor~\eqref{eq:EinsteinTensor}.
We remark that the cosmic strings discussed above constitute a well behaved matter field in our spacetime. 
In fact, the energy-momentum tensor satisfies the conservation law $\nabla \cdot \EneMomTens{\LablCosmicStrings} = 0$
and the dominant energy condition. 

\subsection{%
  \label{subsec:ReductionToSeparatedSystem}
  Reduction to separated systems 
}

We examine {\EinEq} \eqref{eq:EinsteinEquation} by taking
\begin{equation}
   \EneMomTens{} = \EneMomTens{\LablNullDust} + \EneMomTens{\LablCosmicStrings}, 
\end{equation}
where $\EneMomTens{\LablNullDust}$ is given by Eq.~\eqref{eq:EnergyMomentumNullDust} with
$\FourVeloNullDust = \NullFrmFldReeb$ and $\EneMomTens{\LablCosmicStrings}$ is given by
Eq.~\eqref{eq:EnergyMomentumCosmicStrings}. 
Since the Einstein tensor $\EinTens$ is given by Eq.~\eqref{eq:EinsteinTensor},
{\EinEq}~\eqref{eq:EinsteinEquation} yields 
\begin{align}
  \label{eq:EvolutionOfWarpFactor}
  \frac{1}{2a^4} - 2 \frac{a''}{a} = \EinConst \EneDensNullDust,
  \\
  \label{eq:GeometryOfBaseSpace}
  \frac{\BaseRicScal}{2} = \EinConst \mu \NumDens.
\end{align}
These equations show that the roles of the null dust and the cosmic strings are completely separated. 
The null dust, through its energy density $\EneDensNullDust$, determines the ``evolution'' of the warp factor $a(u)$. 
The cosmic strings, through their number density $\NumDens$, determine the geometry of the base space $\BaseSpace$.
Substituting Eq.~\eqref{eq:GeometryOfBaseSpace} to Eq.~\eqref{eq:NonvanishingWeyl}, 
we have a formula for the only {\Nonvanishing} Weyl scalar $\Psi_2$: 
\begin{equation}
   \label{eq:NonVanishingWeylAndString}
   \WeylScal{2} = - \frac{\EinConst \mu \NumDens}{6a^2}. 
\end{equation}

\section{
  \label{sec:GeometryOfBaseSpace}
  Geometry of the base space
}

We study the geometry of the two-dimensional base space $\BaseSpace$ through Eq.~\eqref{eq:GeometryOfBaseSpace},
where the number density $\NumDens$ of the cosmic strings is an arbitrary non-negative function on $\BaseSpace$.

\subsection{%
  \label{subsubsec:Topology}
  Topology
}

We discuss a topological aspect of the base space $\BaseSpace$ under the assumption that $\BaseSpace$ is compact. 
The Einstein equation relates the Euler characteristic of $\BaseSpace$, denoted by $\chi(\BaseSpace)$, and the total number of the cosmic strings, 
$ \TotNumStrings \coloneq \int_{\BaseSpace} \NumDens \sqrt{\det \BaseMet}\,\Dx \Dy $, 
as 
\begin{equation}
   \label{eq:EulerCharacteristicBaseSpace}
   \chi(\BaseSpace) = \frac{\EinConst \mu}{2\pi} \TotNumStrings. 
\end{equation}
This follows from the Gauss-Bonnet-Chern theorem
\begin{align}
  \chi(\BaseSpace) = \frac{1}{4\pi} \int_{\BaseSpace} \BaseRicScal \sqrt{\det \BaseMet} \,\Dx \Dy, 
\end{align}
and Eq.~\eqref{eq:GeometryOfBaseSpace}.

The Euler characteristic of a two-dimensional compact manifold is expressed in terms of the genus $\Genus$ as 
$\chi(\BaseSpace) = 2 - 2 \Genus$.
It immediately follows from Eq.~\eqref{eq:EulerCharacteristicBaseSpace} that
the Euler characteristic $\chi(\BaseSpace)$ can never be negative
because the total number of the cosmic strings, $\TotNumStrings$, is non-negative. 
If there is no cosmic string, i.e., $\TotNumStrings = 0$, 
then $\chi(\BaseSpace)$ vanishes.
On the other hand, if cosmic strings exist, i.e, $\TotNumStrings > 0$, 
then $\chi(\BaseSpace)$ is positive, 
and the only possibility is $\chi(\BaseSpace) = 2$. 
Hence,  the base space $\BaseSpace$ is topologically a sphere. 
In this case, Eq.~\eqref{eq:EulerCharacteristicBaseSpace} yields 
\begin{equation}
   \EinConst \mu = \frac{4 \pi}{\TotNumStrings}. 
\end{equation}

\subsection{
  \label{subsec:Metric}
  Metric%
}

\subsubsection{%
  The case of no cosmic strings
}

We consider the base space metric $\BaseMet$ when there are no cosmic strings. 
In this case, Eq.~\eqref{eq:GeometryOfBaseSpace} leads to a vanishing Ricci scalar, $\BaseRicScal = 0$. 
This means that the two-dimensional base space $\BaseSpace$ is flat. 
Here, we consider the simplest metric functions, $\Omega(x,y) = f(x,y) = 0$, which leads to 
\begin{equation*}
   \BaseMet = \Dx^2 + \Dy^2. 
\end{equation*}
The ranges of the coordinates $x,y$ depend on the topology of the base space $\BaseSpace$.
For example, if $\BaseSpace$ is compact, then $\BaseSpace$ is a flat torus, and thus, the ranges are given by 
\begin{equation}
   \label{eq:CoordinateRanges}
   x \in [0, x_0], \qquad y \in [0, y_0], 
\end{equation}
where $x_0, y_0$ are some positive constants.
For this base space, the three-dimensional K-contact manifold $\ThreeMani$ is obtained by the following identifications: 
\begin{equation}
   (v, x, y + y_0) \sim (v, x, y), \qquad (v, x + x_0, y) \sim (v + x_0 y, x, y). 
\end{equation}
The second identification is necessary for the contact form $\ContForm = \Dv + x \Dy$ to be well defined. 
For this K-contact manifold $\ThreeMani$, the spacetime metric is written as 
\begin{equation}
   \label{eq:MetricWithoutCosmicStrings}
   \FourMet = -2 \Du \qty(\Dv + x \Dy) + a^2(u)\qty(\Dx^2 + \Dy^2). 
\end{equation}
\Rev{This is a so-called conformally flat pure radiation metric, which is  discussed in
  \cite{Edgar:1996dd,Griffiths:1998sp}}. 

\subsubsection{%
  The case of uniformly distributed cosmic strings
}

Let us consider the case that the cosmic strings are uniformly distributed,
i.e., the number density $\NumDens(x,y)$ is a positive constant. 
Then, it follows from Eq.~\eqref{eq:GeometryOfBaseSpace} that $\BaseRicScal$ is also a positive constant.
This implies that $\BaseSpace$ is a round sphere. 
For the metric functions $\Omega(x,y)$ and $f(x,y)$, 
we take the following ansatz:
\begin{equation}
   f(x,y) = 0, \qquad \Omega(x,y) = \Omega(x). 
\end{equation}
Then, Eq.~\eqref{eq:BaseRicSca} gives 
\begin{equation}
   \dv[2]{x} e^{-2\Omega} = - \BaseRicScal. 
\end{equation}
This is readily solved as 
\begin{equation}
   e^{-2\Omega} = \frac{\BaseRicScal}{2} \qty(1 - x^2), 
\end{equation}
Introducing new coordinates $\theta, \phi$ such that
\begin{equation}
   \cos \theta = - x, \qquad \phi = \frac{\BaseRicScal}{2} y,
\end{equation}
we can write the base space metric as 
\begin{equation}
   \BaseMet = \frac{2}{\BaseRicScal} \qty(\dd \theta^2 + \sin^2 \theta \, \dd \phi^2). 
\end{equation}
In the K-contact manifold $\ThreeMani$, we take a new coordinate $\tilde v$ such that
\begin{equation}
   \tilde v = \frac{\BaseRicScal}{2} v. 
\end{equation}
Then, the contact form is written as
\begin{equation}
   \ContForm = \frac{2}{\BaseRicScal} \qty(\dd \tilde v - \cos \theta \dd \phi). 
\end{equation}
Consequently, the spacetime metric is given as 
\begin{equation}
   \FourMet
   =
   \frac{2}{\BaseRicScal}
   \qty[-2 \Du \qty(\dd \tilde v - \cos \theta \dd \phi) + a^2(u) \qty(\dd \theta^2 + \sin^2 \theta \, \dd \phi^2)].
\end{equation}

\section{Evolution of the warp factor}
\label{sec:EvolutionOFWarpFactor}

We investigate the evolution of the warp factor $a(u)$ through Eq.~\eqref{eq:EvolutionOfWarpFactor}. 
This equation shows that the energy density $\EneDensNullDust$ has to be a function only of $u$. 
This ensures that the equation of motion \eqref{eq:ConditionForEnergyDensityOfNullDust} is satisfied.
In the following subsections, we examine three typical models for the energy density $\EneDensNullDust$.

\subsection{The case of no null dust}

We consider the case where the null dust does not exist, i.e., $\EneDensNullDust(u) = 0$. 
In this case, Eq.~\eqref{eq:EvolutionOfWarpFactor} reads
\begin{equation}
  \label{eq:EvolutionOfWarpFactorVacuum}
  \frac{1}{2a^4} - 2 \frac{a''}{a} = 0. 
\end{equation}
This equation is readily solved as
\begin{equation}
   \label{eq:WarpFactorVacuum}
   a(u) = \sqrt{2E u^2 + \frac{1}{8E}},
\end{equation}
where $E$ is a positive constant, 
and we have taken the coordinate $u$ so that $a(u)$ takes the minimum at $u = 0$.

Combining this warp factor with the metric \eqref{eq:MetricWithoutCosmicStrings},
which is derived under the assumption of no cosmic strings, 
we obtain a vacuum metric 
\begin{equation}
   \label{eq:MinkowskiMetricDarbouxCoordinates}
   \FourMet = -2 \Du \qty(\Dv + x \Dy) + \frac{16E^2 u^2 + 1}{8E}\qty(\Dx^2 + \Dy^2). 
\end{equation}
Since the Weyl tensor of the metric \eqref{eq:MetricWithoutCosmicStrings} vanishes,
the solution describes a locally flat spacetime. 
There should exist local coordinates $(U,V,X,Y)$ such that 
\begin{equation}
   \FourMet = - 2 \dd U \dd V + \dd X^2  + \dd Y^2. 
\end{equation} 
The coordinate transformation is explicitly given by 
\begin{align}
  U &= 4Eu,
  \\
  V &= \frac{v}{4E} + \frac{xy}{8E} + \frac{u(x^2 + y^2)}{4},
  \\
  X &= \frac{1}{\sqrt{8E}} \qty(x + 4E u y),
  \\
  Y &= \frac{1}{\sqrt{8E}} \qty(y - 4E u x).    
\end{align}

\subsection{The case of a constant energy density}

\begin{figure}[h]
   \centering
   \includegraphics[]{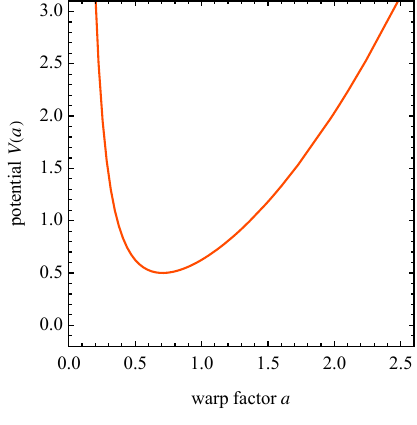}
   \caption{The potential $V(a)$ of Eq.~\eqref{eq:PotentialWithConstEnergyDensity} with $C = 2$.}
   \label{fig:PotentialWithConstEnergyDensity}
\end{figure}

We consider the simple model that the energy density $\EneDensNullDust$ is constant, 
\begin{equation}
   \EneDensNullDust(u) = \EneDensNullDust_0 > 0. 
\end{equation}
In this case, from Eq.~\eqref{eq:EvolutionOfWarpFactor}, we obtain 
\begin{equation}
   \label{eq:IntegralOfEvolutionEquation}
   \frac{1}{2} {a'}^2 + V(a) = E,
\end{equation}
where $E$ is a constant, and $V(a)$ is defined as 
\begin{equation}
   \label{eq:PotentialWithConstEnergyDensity}
   V(a) \coloneq \frac{1}{8a^2} + \frac{C^2}{8} a^2,
   \qquad
   C \coloneq \sqrt{2 \EinConst \EneDensNullDust_0}. 
\end{equation}
We regard Eq.~\eqref{eq:IntegralOfEvolutionEquation} as the energy conservation of a particle moving in one dimension with the 
potential \eqref{eq:PotentialWithConstEnergyDensity}. 
The potential is concave as shown in Fig.~\ref{fig:PotentialWithConstEnergyDensity}. 
Let $V_{\LablMin}$ denote the minimum value of the potential. 
Then, the energy $E$ must satisfy $E \geq V_{\LablMin}$. 
If $E = V_{\LablMin}$, the warp factor is constant. 
Otherwise, the warp factor evolves within the finite region
\begin{equation}
   a_{\LablMin} \leq a \leq a_{\LablMax},    
\end{equation}
where $a_{\LablMin}, a_{\LablMax}$ are the positive roots of the equation $V(a) = E$. 

\subsection{A specific case of an evolving energy density}

\begin{figure}[h]
   \centering
   \includegraphics[]{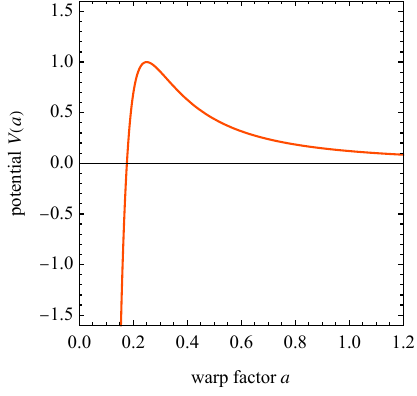}
   \caption{The potential $V(a)$ of Eq.~\eqref{eq:PotentialOfSpecificCase} with 
     $C = 4$. }
   \label{fig:PotentialWithM=-3}
\end{figure}

In the simple model investigated in the previous subsection,
the warp factor is prevented from going to zero by the potential barrier in the vicinity of $a = 0$. 
In contrast, we can also construct a model that allows the warp factor to go to zero and thus the energy density
$\EneDensNullDust(u)$ diverges. 
Let $\EneDensNullDust(u)$ be given by 
\begin{equation}
   \label{eq:AnsatzForEnergyDensityNullDust}
   \EneDensNullDust(u) = \frac{\EneDensNullDust_0}{a^6(u)},
   \qquad
   \EneDensNullDust_0 \neq 0.
\end{equation}
In this case,
Eq.~\eqref{eq:EvolutionOfWarpFactor} is also reduced to the energy conservation~\eqref{eq:IntegralOfEvolutionEquation}, 
where the potential is given by 
\begin{equation}
   \label{eq:PotentialOfSpecificCase}
   V(a) = \frac{1}{8a^2} - \frac{C^2}{16} \frac{1}{a^4},
   \qquad
   C \coloneq \sqrt{2 \EinConst \EneDensNullDust_0}. 
\end{equation}
The shape of the potential is as shown in Fig.~\ref{fig:PotentialWithM=-3}. 
This figure implies that the warp factor is allowed to go to zero,  
and then, the energy density diverges. 

\section{%
\label{sec:Conclusion}
Conclusion}

We have constructed a class of exact solutions of {\EinEq} using a three-dimensional contact manifold with a degenerate metric
that is compatible with the contact structure.

We first extend the notion of metric compatibility, which had only been considered for {\Nondegenerate} metrics, to degenerate ones. 
Then, we construct the spacetime manifold as a product of the real line $\R$ and the three-dimensional contact manifold with a metric 
\begin{equation}
   \tag{\ref{eq:MetricConstruction}}
   \FourMet = - \Du \otimes \ContForm - \ContForm \otimes \Du + a^2 \ContMet, 
\end{equation}
where $\ContForm$ is the contact form, $\ContMet$ is the degenerate contact metric, and $a$ is a function of $u$ called the warp factor. 
We assume that the three-dimensional contact manifold is K-contact, i.e., 
the Reeb vector field $\ReebVect$ is a Killing vector field.

A striking feature of our construction is that the Ricci tensor takes a particularly simple form. 
Indeed, the contravariant Ricci tensor is given by 
\begin{equation}
  \tag{\ref{eq:ContravariantFourRicciTensor}}
  \FourContraRicTens
  =
  \frac{\BaseRicScal}{2a^4} h^{IJ} \FrmFld_I \otimes \FrmFld_J 
  +
  \qty(\frac{1}{2a^4} -2 \frac{a''}{a}) \NullFrmFldReeb \otimes \NullFrmFldReeb.
\end{equation}
This is similar to the Ricci tensors of three-dimensional {\Nondegenerate} K-contact manifolds, which are always $\eta$-Einstein. 
The corresponding property of a three-dimensional degenerate K-contact manifold emerges
when the manifold is realized as a null hypersurface in four-dimensional spacetime constructed by
Eq.~\eqref{eq:MetricConstruction}.

The Ricci tensor~\eqref{eq:ContravariantFourRicciTensor} leads to a class of solutions of {\EinEq}, 
which is worth calling the ``\Emph{contact universe}'',
i.e., the time evolution of a contact manifold. 
The matter fields can be identified with a null dust and cosmic strings. 
The energy density of the null dust and the number density of the cosmic strings can be given arbitrarily.
The roles of the null dust and the cosmic strings are completely separated. 
The energy density of the null dust determines the evolution of the warp factor of the contact universe, 
while the number density of the cosmic strings determines the geometry of the two-dimensional base space $\BaseSpace$. 
For some simple models of these densities, we examined {\EinEq}.

Our spacetime admits two characteristic null vector fields: $\NullFrmFldReeb$ and $\NullFrmFldRL$. 
The null vector field $\NullFrmFldReeb$, which originates from the Reeb vector field, is covariantly constant. 
This implies that the spacetime belongs to the \Rev{Kundt} class \cite{stephani_exact_2003,griffiths2009exact}. 
The other null vector field $\NullFrmFldRL$, which is the metric dual of the contact form, 
generates twisting shear-free null geodesic congruence.
This is the twisting structure characterizing the contact universe. 
The existence of these two null vector fields suggests that the spacetime is algebraically special.
In fact, it is found that the only {\Nonvanishing} Weyl scalar is $\WeylScal{2}$, which is explicitly written as 
\begin{equation}
   \tag{\ref{eq:NonVanishingWeylAndString}}
   \WeylScal{2} = - \frac{\EinConst \mu \NumDens}{6a^2}, 
\end{equation}
where $\NumDens$ is the number density of the cosmic strings.
This indicates that the spacetime is of Petrov type D if the cosmic strings are present,
and of type O (conformally flat) if the strings are absent. 
\Rev{In the latter case, all curvature invariants vanish. }

\begin{acknowledgments}
   HI and YM are partly supported by MEXT Promotion of Distinctive Joint
   Research Center Program JPMXP 0723833165. 
   TK is partly supported by JSPS KAKENHI Grant Number JP20K03772 and 
   MEXT Quantum Leap Flagship Program (MEXT Q-LEAP) Grant Number
   JPMXS0118067285.
\end{acknowledgments}

\appendix

\section{%
  \label{sec:ProofReebIsGeodesic}
  Proof that the Reeb vector field is a geodesic vector field 
}

We show that the Reeb vector field $\ReebVect$ is a geodesic vector field on a contact manifold for all the cases
$\varepsilon = -1, 0, 1$.  
We prove that $\ReebVect$ satisfies the equation, 
\begin{equation}
   \label{eq:EquationGeodesicToBeProved}
   \ContMet(\ContConnect_{\ReebVect}\ReebVect, \X) = 0, 
\end{equation}
where $\X$ is an arbitrary vector field and 
$\ContConnect$ is a torsion-free connection with the metricity.
This connection is uniquely determined for a {\Nondegenerate} metric ($\varepsilon = \pm 1$) and is called the Levi-Civita
connection. On the other hand, this is not the case when the metric is degenerate ($\varepsilon = 0$). 
Equation~\eqref{eq:EquationGeodesicToBeProved} certainly implies that $\ReebVect$ is a geodesic vector field 
\begin{equation}
   \begin{cases}
     \ContConnect_{\ReebVect} \ReebVect
     = 0
     \quad
     \text{if $\ContMet$ is {\Nondegenerate} ($\varepsilon = \pm 1$)}, 
     \\
     \ContConnect_{\ReebVect} \ReebVect
     \propto \ReebVect
     \quad
     \text{if $\ContMet$ is degenerate ($\varepsilon = 0$)}.
   \end{cases}
\end{equation}

In order to prove Eq.~\eqref{eq:EquationGeodesicToBeProved},
we first write the left-hand side of
Eq.~\eqref{eq:EquationGeodesicToBeProved} as follows: 
\begin{equation}
   \ContMet(\ContConnect_{\ReebVect}\ReebVect, \X)
   =
   \ContConnect_{\ReebVect}
   \qty\big[
   \ContMet(\ReebVect, \X)
   ]
   -
   \qty\big(\ContConnect_{\ReebVect} \ContMet)
   (\ReebVect, \X)
   -
   \ContMet(\ReebVect, \ContConnect_{\ReebVect} \X). 
\end{equation}
Then using Eq.~\eqref{eq:MetDualContactFormReeb}
and the metricity of the connection $\ContConnect$, 
we have 
\begin{align}
  \ContMet(\ContConnect_{\ReebVect}\ReebVect, \X)
  &=
    \ContConnect_{\ReebVect}
    \qty\big[
    \varepsilon \ContForm(\X)
    ]
    -
    \ContMet(\ReebVect, \ContConnect_{\ReebVect} \X)
    \notag \\
  \label{eq:IntermideateEquationOfProof}
  &=
  \varepsilon
  \LieDeriv_{\ReebVect}
  \qty\big[
  \ContForm(\X)
  ]
  -
  \ContMet(\ReebVect, \ContConnect_{\ReebVect} \X). 
\end{align}

Next, we rewrite 
$\varepsilon\LieDeriv_{\ReebVect}\qty\big[\ContForm(\X)]$ as follows: 
\begin{align}
  \varepsilon\LieDeriv_{\ReebVect}
  \qty\big[
  \ContForm(\X)
  ]
  &=
    \varepsilon\qty(
    \LieDeriv_{\ReebVect} \ContForm
    )
    \X
    +
    \varepsilon\ContForm (
    \LieDeriv_{\ReebVect} \X
    )
    \notag \\
  &=
    \varepsilon\qty(
    \dd \IntProd_{\ReebVect} \ContForm
    +
    \IntProd_{\ReebVect} \dContForm
    )
    X 
    + 
    \varepsilon\ContForm (\comm{\ReebVect}{\X})
    \notag \\
  &=
    \ContMet(\ReebVect, \comm{\ReebVect}{\X}), 
\end{align}
where we have used Eqs.~\eqref{eq:DefReeb},
\eqref{eq:NormalizationReeb}, 
\eqref{eq:MetDualContactFormReeb},
and \eqref{eq:DegenContactMetric}. 
Substituting this into Eq.~\eqref{eq:IntermideateEquationOfProof}, 
we have 
\begin{align}
  \ContMet(\ContConnect_{\ReebVect}\ReebVect, \X)
  &=
    \ContMet(
    \ReebVect,
    \comm{\ReebVect}{\X} - \ContConnect_{\ReebVect} \X)
    \notag \\
  &=
    -
    \ContMet(
    \ReebVect, \ContConnect_{\X} \ReebVect
    )
    \notag \\
  &=
    -
    \frac{1}{2} \ContConnect_{\X}
    \qty\big[
    \ContMet(\ReebVect, \ReebVect)
    ], 
\end{align}
where we have used the torsion free condition and the metricity.

Finally using $\ContMet(\ReebVect,\ReebVect) = \varepsilon$, 
we obtain Eq.~\eqref{eq:EquationGeodesicToBeProved}.

\section{Sasakian manifold}
\label{sec:SasakianStructure}

We extend the definition of a Sasakian manifold to a degenerate contact manifold.
Then we show that it is equivalent for a contact manifold to be a Sasakian manifold and a K-contact manifold. 

Let $\ContMani$ be a contact manifold with a {\Nondegenerate} metric.
There are some definitions of $\ContMani$ being a Sasakian manifold.
In this paper, we will adopt the definition based on a product manifold $\ProdMani$ with a
\Emph{almost complex structure} $\JTens$ defined by 
\begin{equation}
   \label{eq:DefinitionAlmostComplexStructure}
  \JTens (\X + f \pdv{t}) \coloneqq \PhiTens \X - f \ReebVect + \ContForm(\X) \pdv{t},
\end{equation}
where $\X$ is a vector field tangent to $\ThreeMani$ and $f$ is a function on $\ProdMani$.
If the almost complex structure $\JTens$ is integrable,
i.e., the Nijenhuis torsion $\comm{\JTens}{\JTens}$ vanishes, 
the contact manifold is called the \Emph{Sasakian manifold} \cite{Blair:1976}. 
The Nijenhuis torsion is a tensor field of type  $(1,2)$ given by
\begin{equation}
   \comm{\JTens}{\JTens} (\X',\Y')
   =
   \JTens^2 \comm{\X'}{\Y'} + \comm{\JTens \X'}{\JTens \Y'} - \JTens \comm{\JTens \X'}{\Y'} - \JTens \comm{\X'}{\JTens \Y'}, 
\end{equation}
where $\X'$ and $\Y'$ are vector fields on $\ProdMani$, and the brackets on the right hand side denote the Lie bracket of
vector fields. 
This definition does not use the metric directly. 
Thus, we adopt this definition even for contact manifolds with a degenerate metric. 
The vanishing of the Nijenhuis torsion of $\JTens$ is equivalent to the following condition \cite{Blair:1976}: 
\begin{equation}
   \label{eq:Normality}
   \comm{\PhiTens}{\PhiTens}(X,Y) + \dContForm (X, Y) \ReebVect  = 0.
\end{equation}

In three dimensions, the contact manifold $\ContMani$ is Sasakian if and only if $\ContMani$ is K-contact. 
This can be seen by applying the left hand side of Eq.~\eqref{eq:Normality} to the frame fields
$\ReebVect, \FrmFld_{1}, \FrmFld_{2}$, which are given by Eqs.~\eqref{eq:FrameCoframeDarbouxCoord1}
and \eqref{eq:FrameCoframeDarbouxCoord2}. 
Indeed,  it follows that 
\begin{align}
  \comm{\PhiTens}{\PhiTens}(\ReebVect, \FrmFld_I) + \dContForm(\ReebVect, \FrmFld_I)\ReebVect 
  &= - \ReebVect(\PhiTensComp^J{}_I)\PhiTensComp^{K}{}_{J} \FrmFld_K, 
  \\
  \comm{\PhiTens}{\PhiTens}(\FrmFld_2,\FrmFld_3) + \dContForm(\FrmFld_2,\FrmFld_3)\ReebVect &= 0. 
\end{align}
Thus the contact manifold is Sasakian if and only if $\ReebVect(\PhiTensComp^{I}{}_J) = 0$.
This condition is equivalent to $\ReebVect(\ContMetComp_{IJ}) = 0$, which implies that the Reeb vector field $\ReebVect$ is a
Killing vector field, i.e., the contact manifold is K-contact. 

\bibliography{references}

\end{document}